# Evaluation and selection of models for out-of-sample prediction when the sample size is small relative to the complexity of the data-generating process

HANNES LEEB

*Department of Statistics, Yale University, 24 Hillhouse Avenue, New Haven, CT 06511, USA.*
*E-mail: hannes.leeb@yale.edu*

In regression with random design, we study the problem of selecting a model that performs well for out-of-sample prediction. We do not assume that any of the candidate models under consideration are correct. Our analysis is based on explicit finite-sample results. Our main findings differ from those of other analyses that are based on traditional large-sample limit approximations because we consider a situation where the sample size is small relative to the complexity of the data-generating process, in the sense that the number of parameters in a 'good' model is of the same order as sample size. Also, we allow for the case where the number of candidate models is (much) larger than sample size.

## 1. Introduction

Some of today's most challenging statistical problems feature a large number of potentially important factors or variables and a comparatively small sample size. For example, van't Veer *et al.* (2002) successfully use gene expression profiling to predict recurrence of breast cancer using a classifier comprised of 70 genes that are selected from a total of about 25 000 based on a sample of size 78; see also van de Vijver *et al.* (2002). In such applications, the goal of model selection is often not to find 'the correct model', but rather a model that performs well for prediction. Moreover, the number of explanatory variables (e.g., genes) in the selected model is often of the same order as sample size and the number of candidate models (e.g., subsets of genes) is much larger than sample size. We consider one problem of that kind: regression with random design, where the true model is allowed to be infinite-dimensional, and where the goal is to find a model with 'good' out-of-sample predictive performance.[1] We focus on situations where the sample size is relatively small, in the sense that the number of parameters in a 'good' model is of the same order as sample size. We also allow for the case where the number of candidate models is (much) larger than sample size. To select a 'good' model, we estimate the performance of candidate models and select the one with the best estimated performance.

## 1.1. Classical procedures

Of course, model selection based on estimated predictive performance has already been extensively studied. Methods developed for that aim include the $S_p$ criterion (which can be traced back to Tukey (1967); see also Hocking (1976), Thompson (1976a,1976b)); the Akaike information criterion (AIC; Akaike (1969)); the final prediction error criterion (FPE; Akaike (1970)); the $C_p$ criterion of Mallows (1973); the generalized cross-validation criterion (GCV; Craven and Wahba (1978)); or the small-sample corrected version of AIC (AICc; Hurvich and Tsai (1989)). Minimizing these criteria over a class of candidate models leads to a model selection procedure that is conservative (or over-consistent) in parametric settings.[2] Alternatively, consistent model selection procedures can be used, including the Bayesian information criterion (BIC; Schwarz (1978)) or the minimum description length criterion (MDL; Rissanen (1978)). Further related methods include the prediction criterion (PC) of Amemiya (1980) and the risk inflation criterion (RIC) of Foster and George (1994).

Existing performance analyses of these model selectors do not give a clear picture as to what method is preferable. Consider a so-called post-model-selection estimator, that is, an estimator obtained by first selecting a model based on the training data and then fitting the selected model to the same training data by a method like least-squares or maximum likelihood. In a parametric setting, Kempthorne (1984) showed that any post-model-selection estimator is admissible within the class of all post-model-selection estimators (for squared error in-sample prediction loss). In large samples, it is well known that AIC and similar procedures are asymptotically efficient (in a certain sense) if the true model is infinite-dimensional, while BIC and related methods are efficient if the true model is finite-dimensional (cf. Shao (1997) and the references given therein). In finite samples, however, BIC can be more efficient than AIC (or vice versa), depending on sample size and on the unknown parameters, both in the parametric and the nonparametric cases (cf. Kabaila (1998)). Yang (2005) showed that one cannot find a procedure that combines the strengths of AIC and BIC. In the case where the true model is finite-dimensional and as sample size gets large, consistent model selectors choose the smallest correct model with probability approaching 1, while conservative ones do not; however, consistent model selectors also lead to unbounded worst-case risk, while the worst-case risk corresponding to conservative procedures typically stays bounded in large samples (cf. Leeb and Pötscher (2005) as well as Leeb and Pötscher (2008)). Hence, from the perspective of existing performance analyses, one cannot prefer one of these model selectors over the other in general because the performance of a given model selector depends on unknown parameters and on sample size.

---

[1] Here, 'out-of-sample prediction' means prediction of new responses given hitherto unobserved explanatory variables, whereas 'in-sample prediction' means prediction of new responses for the same explanatory variables as in the training data.

[2] In a parametric setting, most model selectors can be broadly classified as either consistent or conservative: Consistent model selectors are such that the probability of selecting the most parsimonious correct model goes to 1 as sample size increases; conservative model selectors are not consistent, but such that the probability of selecting an incorrect model goes to zero.

## 1.2. New approaches

In this paper, we adopt a different perspective that provides new results and insights. To explain, we first note that the aforementioned analyses that are based on asymptotic considerations rely on large-sample limit approximations that 'kick in' provided that the sample size is 'sufficiently large'; the precise meaning of 'sufficiently large' typically depends on the underlying true data-generating process and more complex processes usually require larger samples.[3] In practice, however, one often faces a very different situation, namely one where the given sample size is relatively small compared to the complexity of the data-generating process, for example, in the sense that the number of parameters in a 'good' model is of the same order as sample size. In addition, the number of candidate models is often (much) larger than sample size. Here, we adopt a framework that is specifically designed for such scenarios.

We find that generalized cross-validation and Tukey's $S_p$ criterion perform well in selecting a 'good' model, even if the candidate models are complex when compared to sample size and also if the number of candidate models is much larger than sample size. More specifically, we show that the true out-of-sample predictive performance of a candidate model is well approximated by generalized cross-validation (or by the objective function of the $S_p$ criterion) with high probability, uniformly over large classes of candidate models and uniformly over huge regions in parameter space under very weak conditions; for details, see Theorem 3.4 and Corollary 3.5. Moreover, we find that several other model selectors, including AIC and BIC, can be systematically defective when evaluating complex models and that their performance can be anything from satisfactory or mildly suboptimal to completely unreasonable, depending on unknown parameters. (This is in stark contrast to the well-known result that generalized cross-validation and the $S_p$ criterion are asymptotically equivalent to AIC – a result that holds asymptotically as the sample size gets large relative to the complexity of the data-generating process.) Our findings are based on explicit finite-sample results (cf. Theorem 3.2 and Corollary 3.3) and backed up by simulation examples.

---

[3] Here, the precise meaning of 'complexity' depends on the details of the approximation that is being considered. In many cases, 'complexity' is related to the number of parameters or to smoothness conditions.

Conceptually, our approach is inspired by Beran (1996), Beran and Dümbgen (1998) and Beran (2000). The setting in these papers is similar to ours, in that the number of explanatory variables is of the same order as sample size. However, these papers consider regression with fixed design and the focus is on estimating a different performance measure, namely on the Euclidean distance between the true location parameter and the estimate. In that setting, the performance of any estimator depends only on the estimator itself and on the unknown true regression parameter. In contrast, we focus on the out-of-sample predictive performance and we consider random design. In our setting, the out-of-sample predictive performance of any estimator, in addition to depending on the estimator itself and the regression parameter, also depends on the design distribution, which is unknown. (If the number of design variables under consideration is sufficiently small in relation to sample size, the empirical distribution of these design variables can be used as a proxy for the true design distribution. In the setting that we consider, however, this does not work because the number of design variables considered is not necessarily small relative to sample size.) In the setting of Beran (1996), a $C_p$- or AIC-like approach to loss estimation is shown to work well. In our setting, we find that $C_p$ and AIC do not work well and that a different approach to performance estimation is required.

A related direction of research was initiated by Barron, Birgé and Massart (1999) and further explored by Yang (1999) and Baraud (2002); see also Wegkamp (2003), as well as the references in these papers. Instead of attempting to estimate the performance of candidate models, these papers provide finite-sample upper bounds for the risk of post-model-selection estimators that are based on minimizing an objective function like penalized maximum likelihood or penalized least-squares, where the risk is defined as the expected Euclidean distance between the true regression parameter and the estimator, or some similar (known) distance measures, as in Baraud (2002). Under some conditions and for $C_p$- or AIC-like penalty functions, these upper bounds give so-called oracle inequalities, stating that the true risk of the post-model-selection estimator is within a constant multiple of the risk obtained by fitting the minimal-risk model. (Note that the upper bound provided by such an oracle inequality is not known in practice because it depends on the unknown regression parameter.) Our results differ from these in two important aspects. First, we consider a different objective, namely minimizing the out-of-sample prediction risk (where the performance of an estimator, in addition to depending on the estimator and on the true regression parameter, also depends on the unknown distribution of the design variables), and we focus on the case where the sample size is small relative to the complexity of the data-generating process, a case where $C_p$- or AIC-like objective functions do not perform well. Second, instead of giving upper bounds, we show that the performance of the resulting post-model-selection estimator can actually be estimated in our setting. We give finite-sample bounds on the estimation error probability that depend only on quantities that are either known or that can be estimated in a uniformly consistent fashion.

Technically, our paper relies heavily on the results of Breiman and Freedman (1983), who give a large-sample limit analysis of model selection by the $S_p$ criterion for the special case of nested candidate models. A precursor version of our paper that was written in 2005 was instead based on the Marčenko–Pastur law (cf. Marčenko and Pastur (1967)).

### 1.3. Outline of this paper

For a sample of $n$ observations from some data-generating process to be specified later, we consider a collection $\mathcal{M}_n$ of candidate models $m \in \mathcal{M}_n$ with dimension $|m|$. (We use the symbols $m$ to denote a candidate model and $|m|$ to denote the number of explanatory variables in the model $m$.) We do not assume that the true regression function is correctly described or even well approximated by any of the candidate models. Under model $m$, the response is related to a collection of $|m|$ explanatory variables. The leading case of interest is where the sample size is small relative to the complexity of the true data-generating process, a case where 'interesting' models are such that $|m|/n$ is large, for example, $|m|/n$ equals 0.1, 0.5 or even 0.9. We focus on the case where the number of candidate models, that is, $\#\mathcal{M}_n$, is as large as, or much larger than, sample size.[4] Our objective is to select a model that performs well for out-of-sample prediction, that is, for predicting a new response given hitherto unobserved explanatory variables. For a fixed set of new explanatory variables, the model that performs best for predicting the corresponding response can, and typically will, depend on the values of these explanatory variables (cf. Claeskens and Hjort (2003)). To identify a model that performs well in an overall sense, we consider random design and we evaluate a model's performance by the conditional mean squared error of the corresponding predictor, where the conditioning is on the training sample. In other words, we search for a model that, when fitted to the given training sample, performs well on average when repeatedly predicting new responses. (Of course, the case of random design also is a scenario of interest in its own right.) The conditional out-of-sample prediction error associated with model $m$ is denoted by $\rho^2(m)$ and we consider the generalized cross-validation criterion GCV($m$) and Tukey's $S_p$ criterion $S_p(m)$, as well as an auxiliary criterion $\hat{\rho}^2(m)$ (that will be defined later), as estimators for $\rho^2(m)$; see Section 2 for the details. We also consider other model selection criteria, namely the Akaike information criterion AIC($m$), Hurvich and Tsai's AICc($m$) and the final prediction error criterion FPE($m$), as well as the Bayesian information criterion BIC($m$).

A theoretical analysis of the aforementioned criteria is given in Section 3, under the assumption that the data are sampled from a Gaussian distribution. We first give an explicit finite-sample analysis of the auxiliary criterion $\hat{\rho}^2(m)$ in Section 3.1. These results allow us to show that generalized cross-validation and the $S_p$ criterion can be used to select a good model with high probability, uniformly over large families of candidate models and uniformly over huge regions in parameter space under very weak conditions; see Section 3.2. Finally, the performance of other model selectors is discussed in Section 3.3. (On a technical level, the results in Section 3 rely heavily on the assumption of Gaussianity, but we suspect that similar findings might be obtained in more general settings and our simulation results appear to support this.)

---

[4]Huber (1973) considers a related setting, where the dimension of the overall model, denoted by $k$, is finite, but increases with $n$ such that $k/n \to 0$. He notes that settings where $k/n$ and $|m|/n$ are large "are unlikely to yield a reasonably simple asymptotic theory" (cf. page 802 of that paper). See also Portnoy (1984, 1985) and Mammen (1989).

The impact of our theoretical results is demonstrated in a simulation study in Section 4, where we also consider non-Gaussian samples. Our simulations include examples where a sample of size 1 300 is used to select a model from over a million candidate models. We demonstrate that model selection by generalized cross-validation or the $S_p$ criterion performs very well here and that the performance of these model selectors is basically unaffected by departures from normality. For the other model selectors that we consider, that is, for $AIC(m)$, $AICc(m)$, $FPE(m)$ and $BIC(m)$, we find in the theoretical analysis and in the simulation examples that their performance can be anything from satisfactory or mildly suboptimal to completely unreasonable, depending on unknown parameters. The more technical parts of the proofs are given in the Appendix.

## 2. Setting of the analysis

Consider a response $y$ that is related to explanatory variables $x = (x_j)_{j=1}^{\infty}$ by

$$y = \sum_{j=1}^{\infty} x_j \beta_j + u \tag{1}$$

for some $\beta = (\beta_j)_{j=1}^{\infty}$. Throughout, we assume that the error $u$ has mean zero and variance $\sigma^2 \geq 0$, and that the (stochastic) sequence of explanatory variables $x = (x_j)_{j=1}^{\infty}$ has mean zero and variance/covariance net $\Sigma = [E(x_i x_j)]_{i,j \geq 1}$ such that the series in (1) converges in squared mean. Moreover, we also assume that the explanatory variables $x_j$, $j \geq 1$, are each uncorrelated with the error $u$ and that the $x_j$'s are not perfectly correlated among themselves.[5] The unknown parameters here are the sequence of regression coefficients, the error variance and the variance/covariance net of the regressors, that is, $\beta$, $\sigma^2$ and $\Sigma$. The (minimal) requirement, that the series in (1) converges in squared mean, restricts $\beta$ in a way that depends on $\Sigma$. For example, if $\Sigma$ is such that the $x_j$'s have variance 1 and are uncorrelated with each other, then it is required that $\beta \in l_2$, that is, $\sum_j \beta_j^2 < \infty$. (For the case where the explanatory variables are not centered, extensions of the results in this paper are given by Leeb (2007).)

Consider a sample of size $n$ from (1). The sample will be denoted by $(Y, X)$, where $Y$ is the $n$-vector $Y = (y^{(1)}, \ldots, y^{(n)})'$, $X$ is the $n \times \infty$ net $X = (x^{(1)'}, \ldots, x^{(n)'})'$ and $(y^{(i)}, x^{(i)})$ are independent and identically distributed copies of $(y, x)$, as in (1). Let $P_{n,\beta,\sigma,\Sigma}$ denote the distribution of the sample $(Y, X)$ and let $E_{n,\beta,\sigma,\Sigma}$ denote the corresponding expectation operator. Similarly, we write $\text{Var}_{\beta,\sigma,\Sigma}[y]$ for the variance of $y$ in (1).[6]

As estimators for $\beta$, we consider restricted least-squares estimators corresponding to submodels of the overall model (1), under which some coefficients of $\beta$ are restricted to zero. Each such submodel can be identified by a 0-1 sequence $m = (m_j)_{j=1}^{\infty}$, where $m_j = 0$ if the $j$th coefficient of $\beta$ is restricted to zero and $m_j = 1$ otherwise; the number of unrestricted components, that is, the number of 1's in $m$, is denoted by $|m|$. Throughout the paper, we shall always assume that $|m| < n - 1$.[7] We call $|m|$ the order of the model $m$. The restricted least-squares estimator corresponding to the model $m$ is denoted by $\tilde{\beta}(m)$ and is defined as follows: $\tilde{\beta}(m)$ is such that its $j$th component equals zero whenever $m_j = 0$; the $|m|$ remaining (unrestricted) components of $\tilde{\beta}(m)$ are obtained by regressing $Y$ on the corresponding columns of $X$.

---

[5]In other words, we require, for each $k \geq 1$ and integers $j_1 < j_2 < \cdots < j_k$, that $(x_{j_1}, \ldots, x_{j_k})'$ is a random vector with mean zero and positive definite variance/covariance matrix that is uncorrelated with $u$.

[6]It should be noted that $P_{n,\beta,\sigma,\Sigma}$, $E_{n,\beta,\sigma,\Sigma}$ and $\text{Var}_{\beta,\sigma,\Sigma}[y]$, in addition to depending on the parameters $\beta$, $\sigma$ and $\Sigma$, also depend on the actual distribution of $(y, x)$ in (1); this dependence is not reflected explicitly by the notation. The distribution of $(y, x)$ will always be clear from the context.

Based on the sample $(Y, X)$, our objective is to find a 'good' model for out-of-sample prediction. To this end, consider a new copy $(y^{(f)}, x^{(f)})$ of $(y, x)$, as in (1), that is independent of $(Y, X)$. Given a model $m$ with $|m| < n - 1$ and the corresponding restricted least-squares estimator $\tilde{\beta}(m)$, we will use $x^{(f)'}\tilde{\beta}(m)$ as a predictor for $y^{(f)}$. To evaluate the performance of this predictor, we consider the conditional and unconditional mean squared prediction errors given by

$$\rho^2(m) = E_{n,\beta,\sigma,\Sigma}[(y^{(f)} - x^{(f)'}\tilde{\beta}(m))^2 \| Y, X]$$

and

$$R^2(m) = E_{n,\beta,\sigma,\Sigma}[(y^{(f)} - x^{(f)'}\tilde{\beta}(m))^2],$$

respectively. For the conditional mean squared prediction error $\rho^2(m)$, note that the sample $(Y, X)$ is kept fixed and the average is taken only with respect to $(y^{(f)}, x^{(f)})$, so $\rho^2(m)$ is a function of $\tilde{\beta}(m) - \beta$. In particular, $\rho^2(m)$ can become large if either the model is too complex (so that $\tilde{\beta}(m)$ is not close to $\beta$ because of over-fit), or if important explanatory variables are not included in the model (so that $\tilde{\beta}(m)$ is not close to $\beta$ because of under-fit). Also, note that $\rho^2(m) = \text{Var}_{n,\theta,\sigma,\Sigma}[y^{(f)} - x^{(f)'}\tilde{\beta}(m) \| Y, X]$ here because the mean of $x^{(f)}$ is zero. For the case where the mean of $x^{(f)}$ is not zero such that $\rho^2(m) = \text{Var}_{n,\theta,\sigma,\Sigma}[y^{(f)} - x^{(f)'}\tilde{\beta}(m) \| Y, X] + (E_{n,\theta,\sigma,\Sigma}[y^{(f)} - x^{(f)'}\tilde{\beta}(m) \| Y, X])^2$, we refer to Leeb (2007): assuming that the sample is Gaussian and that the model includes an intercept, it is shown in that paper that the squared bias, that is, $(E_{n,\theta,\sigma,\Sigma}[y^{(f)} - x^{(f)'}\tilde{\beta}(m) | Y, X])^2$, is of smaller order than the variance, that is, $\text{Var}_{n,\theta,\sigma,\Sigma}[y^{(f)} - x^{(f)'}\tilde{\beta}(m) \| Y, X]$. Our main focus will be on the conditional mean squared prediction error, that is, $\rho^2(m)$, rather than on the unconditional mean squared prediction error, that is, $R^2(m)$, which is based on averaging over hypothetical samples. Also, note that $\rho^2(m)$ depends only on $n$, $\beta$, $\sigma$ and $\Sigma$; $R^2(m)$, on the other hand, also depends on the actual distribution of the random variables in (1).

**Remark 2.1.** Instead of $R^2(m)$ or $\rho^2(m)$, traditional large-sample limit analyses often consider error measures like the (unconditional) mean of $(x^{(f)'}\beta - x^{(f)'}\tilde{\beta}(m))^2$, scaled by a parametric or nonparametric rate (depending on the setting). In a parametric setting, where the parameter $\beta$ has only finitely many non-zero components, this is because the mean of $(x^{(f)'}\beta - x^{(f)'}\tilde{\beta}(m))^2$ goes to zero at a rate of $1/n$, provided that the model $m$ is correct. Similar considerations apply in nonparametric settings under smoothness conditions, provided that the dimension of the model increases appropriately with sample size. Considering the mean of $(x^{(f)'}\beta - x^{(f)'}\tilde{\beta}(m))^2$ or of $(y^{(f)} - x^{(f)'}\tilde{\beta}(m))^2$ is equivalent, as far as selecting a 'good' model is concerned, because the two means differ by a fixed constant, namely the error variance $\sigma^2$. The lack of scaling by some rate in $\rho^2(m)$ and $R^2(m)$ is caused by the fact that we do not assume a parametric model and we do not impose smoothness conditions in a nonparametric model because such assumptions would mean that estimation errors go to zero as sample size increases. Instead, we use approximations that retain the finite-sample feature that estimation errors are potentially large because the sample size is small relative to the complexity of the data-generating process.

---

[7] We assume $|m| < n - 1$ for the sake of convenience; some of our results also hold for $|m| < n$, while others even hold for $|m| \le n$.

The conditional and unconditional mean squared prediction errors depend on unknown parameters and thus must be estimated. For a candidate model $m$, we consider the generalized cross-validation criterion $\mathrm{GCV}(m)$, the $S_p$ criterion $S_p(m)$ and an auxiliary criterion $\hat{\rho}^2(m)$, which are defined as follows. Let

$$\mathrm{GCV}(m) = \frac{(1/n)\mathrm{RSS}(m)}{(1-|m|/n)^2} = \frac{\mathrm{RSS}(m)}{n-|m|}\frac{n}{n-|m|}.$$

In the above display, $\mathrm{RSS}(m)$ denotes the residual sum of squares obtained by fitting the model $m$, that is, $\mathrm{RSS}(m) = \sum_{i=1}^{n}(y^{(i)} - x^{(i)'}\hat{\beta}(m))^2$. The generalized cross-validation criterion is closely related to the $S_p$ criterion, which is defined by

$$S_p(m) = \frac{\mathrm{RSS}(m)}{n-|m|}\frac{n-1}{n-1-|m|}.$$

For technical reasons, we also consider another quantity that is closely related to both $\mathrm{GCV}(m)$ and $S_p(m)$, namely

$$\hat{\rho}^2(m) = \frac{\mathrm{RSS}(m)}{n-|m|}\frac{n+1}{n+1-|m|}.$$

For most practical purposes, the difference between $\mathrm{GCV}(m)$, $S_p(m)$ and $\hat{\rho}^2(m)$ is negligible. (Also, note that $\mathrm{GCV}(m)$, $S_p(m)$ and $\hat{\rho}^2(m)$ are well defined because we always assume that $|m| < n-1$.) The other model selectors mentioned in the Introduction are defined in Section 3.3.

## 3. Theoretical analysis

In this section, we study the problem of estimating the conditional and unconditional mean squared prediction error in the case where the sample is drawn from a Gaussian distribution. We hence assume throughout this section that the random variables in (1) are jointly normal.[8] Unless otherwise noted, we fix parameters $\beta$, $\sigma$ and $\Sigma$ as in (1) and consider a fixed sample size $n$ and a fixed model $m$ with $|m| < n - 1$. For $y$ and $x$ as in (1), set

$$\sigma^2(m) = \mathrm{Var}_{\beta,\sigma,\Sigma}[y\|x_j : j \in \mathbb{N}, m_j = 1].$$

Note that $\sigma^2(m)$ is non-random because the involved random variables are jointly Gaussian, and that $\sigma^2(m) \leq \sigma^2(0) = \mathrm{Var}_{\beta,\sigma,\Sigma}[y]$.

## 3.1. Finite-sample results

The following result (whose first statement is adapted from Breiman and Freedman (1983)) provides the basis for a finite-sample analysis.

**Proposition 3.1.** (i) *The conditional mean squared prediction error $\rho^2(m)$ has the same distribution as 1 plus the ratio of two independent chi-squared random variables with $|m|$ and $n - |m| + 1$ degrees of freedom, respectively, multiplied by $\sigma^2(m)$:*

$$\rho^2(m) \sim \sigma^2(m)\left(1 + \frac{\chi^2_{|m|}}{\chi^2_{n-|m|+1}}\right).$$

(ii) *The residual sum of squares has the same distribution as a chi-squared random variable with $n - |m|$ degrees of freedom, multiplied by $\sigma^2(m)$:*

$$\mathrm{RSS}(m) \sim \sigma^2(m)\chi^2_{n-|m|}.$$

Proposition 3.1 immediately implies that the unconditional mean squared prediction error $R^2(m)$ can be computed explicitly as

$$R^2(m) = \sigma^2(m)\frac{n-1}{n-1-|m|}$$

because we always assume that $|m| < n - 1$ (recall the formula for the mean of the $F$-distribution). This also gives the well-known result that $S_p(m)$ is an unbiased estimator for the unconditional mean squared prediction error $R^2(m)$; the estimators $\mathrm{GCV}(m)$ and $\hat{\rho}^2(m)$ for $R^2(m)$ are biased, but the bias is typically negligible. For $|m| < n - 3$, we also get that the variance of $\rho^2(m)$ is finite and given by

$$2\sigma^4(m)\frac{|m|(n-1)}{(n-|m|-1)^2(n-|m|-3)} \approx \frac{2}{n}\sigma^4(m)\frac{|m|/n}{(1-|m|/n)^3}.$$

We see that the conditional mean squared prediction error, that is, $\rho^2(m)$, is highly concentrated around its mean, that is, $R^2(m)$, provided only that $n$ is large enough relative to $\sigma^4(m)/(1 - |m|/n)^3$. This suggests that $S_p(m)$, $\mathrm{GCV}(m)$ and $\hat{\rho}^2(m)$ can be used to estimate not only $R^2(m)$, but also the conditional mean squared prediction error $\rho^2(m)$. In order to use these considerations for model selection, we need to establish that, say, $\hat{\rho}^2(m)$ is close to $\rho^2(m)$ with high probability, not only for a fixed model $m$, but for an entire collection of candidate models. This is accomplished by the following theorem and the attending corollary.

---

[8]Note that assuming the sample to be Gaussian entails that $P_{n,\beta,\sigma,\Sigma}$ and $E_{n,\beta,\sigma,\Sigma}$, as well as $\mathrm{Var}_{\beta,\sigma,\Sigma}[y]$, are uniquely determined by the parameters in the subscript.

**Theorem 3.2.** *For each* $\epsilon > 0$, *we have*

$$P_{n,\beta,\sigma,\Sigma}(|\hat{\rho}^2(m) - \rho^2(m)| > \epsilon)$$
$$\leq 4 \exp\left[-n\left(1 - \frac{|m|}{n}\right)\Psi\left(\frac{\epsilon}{2\sigma^2(m)}\left(1 - \frac{|m|}{n}\right)\right)\right],$$

*where* $\Psi(\cdot)$ *is defined by* $\Psi(x) = (x/(x+1))^2/8$ *for* $x \geq 0$. *(In the case* $\sigma^2(m) = 0$, *the upper bound is to be interpreted as zero.)*

For fixed $\epsilon > 0$, the upper bound in Theorem 3.2 is of the form $4\exp[-nC]$, where $C$ is always positive. This upper bound is exponentially small in $n$, provided only that $|m|/n$ is bounded away from 1 and $\sigma^2(m)$ is bounded away from infinity. The upper bound depends on the known quantities $n$, $|m|/n$ and $\epsilon$, and also on $\sigma^2(m)$, which is unknown. However, recall that $\sigma^2(m)$ is bounded from above by $\sigma^2(0) = \mathrm{Var}_{\beta,\sigma,\Sigma}[y]$, that is, by the variance of $y$ in (1), which can be readily estimated from the sample, for example, by $(n-1)^{-1}\sum_{i=1}^{n}(y^{(i)} - \bar{y})^2$, where $\bar{y}$ denotes the mean of the responses $y^{(1)}, \ldots, y^{(n)}$ in the training sample. Thus, we see that the upper bound in Theorem 3.2 is exponentially small in $n$, provided only that both the complexity of the candidate model and the variance of the response, that is, $|m|$ and $\mathrm{Var}_{\beta,\sigma,\Sigma}[y]$, are not too large. These considerations, together with Bonferroni's inequality, immediately lead to the following result.

**Corollary 3.3.** *Consider a (finite and non-empty) collection* $\mathcal{M}_n$ *of candidate models and let* $r_n = \sup_{m \in \mathcal{M}_n} |m|/n$. *Then*

$$\sup_{\substack{\beta,\sigma,\Sigma \text{ as in } (1) \\ \mathrm{Var}_{\beta,\sigma,\Sigma}[y] \leq c}} P_{n,\beta,\sigma,\Sigma}\left(\sup_{m \in \mathcal{M}_n} |\hat{\rho}^2(m) - \rho^2(m)| > \epsilon\right)$$
$$\leq 4 \#\mathcal{M}_n \exp[-n(1 - r_n)\Psi((\epsilon/(2c))(1 - r_n))]$$

*for each* $\epsilon > 0$ *and for each (finite)* $c > 0$. *(Here,* $\#\mathcal{M}_n$ *denotes the number of candidate models and* $\Psi(\cdot)$ *is as in Theorem 3.2.)*

The upper bound given in Corollary 3.3 is of the form $4\exp[-nD + \log \#\mathcal{M}_n]$, where the constant $D > 0$ depends on $r_n$ and $c$ (for fixed $\epsilon > 0$). In particular, the upper bound is small provided only that the variance of the response, the complexity of the candidate models and the number of candidate models are not too large in relation to sample size.

Clearly, results paralleling Theorem 3.2 and Corollary 3.3 can also be derived when either generalized cross-validation or the $S_p$ criterion, that is, GCV($\cdot$) or $S_p(\cdot)$, are used instead of $\hat{\rho}^2(\cdot)$. The reason for considering $\hat{\rho}^2(\cdot)$ here is that this estimator leads to the most simple and most revealing upper bound. (In most practical cases, the distinction between $\hat{\rho}^2(m)$, GCV($m$) and $S_p(m)$ is negligible anyway. In the Appendix, we also give a variant of Theorem 3.2 with $S_p(m)$ and $R^2(m)$ replacing $\hat{\rho}^2(m)$ and $\rho^2(m)$, respectively; see Proposition A.5.)

It should be noted that the upper bound in Theorem 3.2 does not go to zero as $\epsilon$ goes to infinity. (The same also applies to the upper bound in Corollary 3.3, which is derived from that in Theorem 3.2.) The upper bound in Theorem 3.2 is, in fact, based on a tighter, but more complicated, bound that is given in Proposition A.4 in the Appendix; that tighter bound does go to zero as $\epsilon \to \infty$. We present the bound of Theorem 3.2 as our main result because, for fixed $\epsilon > 0$, it captures in a simple expression the essential interplay between the sample size, the complexity of the candidate model and the data-generating process that guarantees that the probability of $|\hat{\rho}^2(m) - \rho^2(m)|$ exceeding $\epsilon$ is exponentially small in $n$. The tighter bound of Proposition A.4 does the same, but is much more complicated. Moreover, the upper bound in Theorem 3.2 is tight enough to give the rates of convergence that are presented in the following section.

## 3.2. Approximation results

In this section, we provide conditions under which the upper bounds given previously go to zero. Under these conditions, $\hat{\rho}^2(m)$, GCV($m$) and $S_p(m)$ are close to $\rho^2(m)$ with probability approaching one, uniformly over a collection of candidate models and uniformly over a large region in parameter space. For the sake of simplicity, the results that follow simply state that estimation errors go to zero in probability at a certain rate, instead of giving explicit, but more complicated, finite-sample upper bounds.

**Theorem 3.4.** *For each sample size* $n$, *consider a (finite and non-empty) family* $\mathcal{M}_n$ *of candidate models, let* $r_n = \sup_{m \in \mathcal{M}_n} |m|/n$ *and define* $a_n$ *as*

$$a_n = \sqrt{\frac{\log(\#\mathcal{M}_n + 1)}{n(1 - r_n)^3}}. \tag{2}$$

*Assume that* $a_n \to 0$ *as* $n \to \infty$. *Then*

$$\sup_{m \in \mathcal{M}_n} |\text{GCV}(m) - \rho^2(m)| = O_p(a_n) \tag{3}$$

*holds uniformly over all data-generating processes as in (1) that satisfy* $\text{Var}_{\beta, \sigma, \Sigma}[y] \leq c$ *(where* $c > 0$ *is an arbitrary fixed (finite) constant). Hence, over the indicated set of parameters,* GCV($m$) *is a uniformly* $1/a_n$-*consistent estimator for* $\rho^2(m)$, *uniformly in* $m \in \mathcal{M}_n$. *The same applies with* $S_p(m)$ *or* $\hat{\rho}^2(m)$ *in place of* GCV($m$). *(These statements all continue to be true if* $R^2(m)$ *replaces* $\rho^2(m)$.)*

Informally, the condition that $a_n \to 0$ maintained by Theorem 3.4 imposes two requirements on the complexity of the candidate models and on the number of candidate models, respectively, in relation to sample size: (i) that the candidate models are not too complex, that is, $r_n$ is not too close to 1, so that $n(1 - r_n)^3$ can get large; (ii) that the number of candidate models is not too large, in the sense that $\log \# \mathcal{M}_n$ is of smaller order than $n(1 - r_n)^3$. The first requirement, that is, that $r_n$ is not too close to 1, only rules out cases that are susceptible to severe over-fit anyway. The second requirement, that is, that $\log \# \mathcal{M}_n = o(n(1 - r_n)^3)$, rules out certain cases of complete subset selection, for example, the case where $\# \mathcal{M}_n = 2^n$. However, that requirement typically still allows for considerably large classes of candidate models. In practice, limited computational resources will often entail much stronger restrictions on the number of candidate models that can be considered. The consequences of Theorem 3.4 for model selection are immediate.

**Corollary 3.5.** *In the setting of Theorem 3.4, assume that $a_n \to 0$. Consider (measurable) minimizers of* $\mathrm{GCV}(m)$ *and* $\hat{\rho}^2(m)$ *over* $\mathcal{M}_n$,

$$\hat{m}_n^* = \underset{m \in \mathcal{M}_n}{\arg\min} \, \mathrm{GCV}(m) \quad and \quad m_n^* = \underset{m \in \mathcal{M}_n}{\arg\min} \, \rho^2(m).$$

(i) *The empirically best model, that is, $\hat{m}_n^*$, is asymptotically as good as the best model, in the sense that*

$$|\rho^2(\hat{m}_n^*) - \rho^2(m_n^*)| = O_p(a_n),$$

*uniformly over all data-generating processes as in (1) that satisfy* $\mathrm{Var}_{\beta, \sigma, \Sigma}[y] \le c$ *(where $c > 0$ is an arbitrary fixed (finite) constant).*

(ii) *The predictive performance of the model $\hat{m}_n^*$ can be estimated in a uniformly consistent fashion, in the sense that*

$$|\mathrm{GCV}(\hat{m}_n^*) - \rho^2(\hat{m}_n^*)| = O_p(a_n),$$

*uniformly over all data-generating processes as in (1) that satisfy* $\mathrm{Var}_{\beta, \sigma, \Sigma}[y] \le c$.

*The above continues to hold if, throughout, $\mathrm{GCV}(\cdot)$ is replaced by $S_p(\cdot)$ or $\hat{\rho}^2(\cdot)$. (These statements continue to be true if $R^2(\cdot)$ replaces $\rho^2(\cdot)$.)*

If Corollary 3.5 applies, the generalized cross-validation criterion (or, equivalently, either the $S_p$ criterion or $\hat{\rho}^2(m)$) can be used to select a good model whose estimated performance is close to its actual performance (with probability approaching 1), uniformly over the indicated region in parameter space. That region in parameter space is characterized by an upper bound on $\mathrm{Var}_{\beta, \sigma, \Sigma}[y]$, that is, on the variance of the response $y$ in the overall model (1). Boundedness of the response's variance is a very innocuous restriction, showing that the performance of generalized cross-validation (or the related criteria $S_p(m)$ and $\hat{\rho}^2(m)$) is guaranteed over a huge region in parameter space: for example, fix $\sigma^2$ and fix $\Sigma$ such that the explanatory variables in (1) are uncorrelated with unit variance; for $c > \sigma^2$, the condition $\text{Var}_{\beta,\sigma,\Sigma}[y] \leq c$ then requires that $\beta$ satisfies $\sum_j \beta_j^2 \leq c - \sigma^2$, that is, $\beta$ can range over a non-compact subset of $l_2$.

For parameters $\beta$, $\sigma$ and $\Sigma$ satisfying $\text{Var}_{\beta,\sigma,\Sigma}[y] \leq c$, for a given model $m$ and for a fixed sample size $n$, the conditional and unconditional mean squared prediction error can take on any value between between $\sigma^2$ and $\text{Var}_{\beta,\sigma,\Sigma}[y]$ (because the model $m$ can contain anything between all and none of the non-zero coefficients of $\beta$). By considering such parameters in Corollary 3.3, Theorem 3.4 and Corollary 3.5, we focus on situations where the noise $u^{(f)}$ is not the dominant source of error when predicting $y^{(f)} = x^{(f)'}\beta + u^{(f)}$ out-of-sample. This captures scenarios where the sample size is small, relative to the complexity of the true data-generating process. Our results show that generalized cross-validation (or the $S_p$ criterion or $\hat{\rho}^2(m)$) performs very well in such situations.

## 3.3. Other model selectors

It is instructive to compare generalized cross-validation and the $S_p$ criterion to other model selection methods. We consider some classical examples, namely the Akaike information criterion (AIC), the AIC with finite-sample correction (AICc) of Hurvich and Tsai, Akaike's final prediction error criterion (FPE), and Schwarz' Bayesian information criterion (BIC), whose objective functions are given by $\text{AIC}(m) = n^{-1}\text{RSS}(m)\exp(2|m|/n)$, $\text{AICc}(m) = n^{-1}\text{RSS}(m)\exp(2(|m|+1)/(n-|m|-2))$, $\text{FPE}(m) = n^{-1}\text{RSS}(m)(1 + |m|/n)/(1 - |m|/n)$ and $\text{BIC}(m) = n^{-1}\text{RSS}(m)n^{|m|/n}$, respectively. (Traditionally, AIC, AICc and BIC are defined on a logarithmic scale; the equivalent exponential scale used here is more convenient for our purposes. We also assume here that $|m| < n - 2$, to ensure that $\text{AICc}(m)$ is well defined.) Note that the objective functions of AIC, AICc, FPE and BIC are strictly increasing in $\text{RSS}(m)$ and that the same is true for $\text{GCV}(m)$. This allows us to express, say, $\text{AIC}(m)$ as $\text{AIC}(m) = \text{GCV}(m)e^{2|m|/n}(1 - |m|/n)^2$, informally suggesting the following: The AIC-objective function $\text{AIC}(m)$ is close to

$$\rho^2(m)e^{2|m|/n}(1 - |m|/n)^2; \tag{4}$$

the FPE-objective function $\text{FPE}(m)$ is close to

$$\rho^2(m)(1 + |m|/n)(1 - |m|/n); \tag{5}$$

the objective function $\text{AICc}(m)$ is close to

$$\rho^2(m)e^{2(|m|+1)/(n-|m|-2)}(1 - |m|/n)^2; \tag{6}$$

and $\text{BIC}(m)$ is close to

$$\rho^2(m)n^{|m|/n}(1 - |m|/n)^2. \tag{7}$$

More formally, in the setting of Theorem 3.4 and provided that the quantity $a_n$ defined there goes to zero, the differences between $\text{AIC}(m)$ and $\text{FPE}(m)$ and the quantities in

(4) and (5), respectively, converge to zero uniformly in $m \in \mathcal{M}_n$, where convergence is in probability, uniformly over the set of parameters satisfying $\mathrm{Var}_{\beta,\sigma,\Sigma}[y] \leq c$ with $c > 0$. (For AIC, this immediately follows from Theorem 3.4 because the events where $|\mathrm{AIC}(m) - \rho^2(m)e^{2|m|/n}(1 - |m|/n)^2| > \epsilon$ and where $|\mathrm{GCV}(m) - \rho^2(m)| > \epsilon e^{-2|m|/n}(1 - |m|/n)^{-2}$ coincide and are contained in the event where $|\mathrm{GCV}(m) - \rho^2(m)| > \epsilon e^{-2}$; for FPE, a similar argument applies.) The same is true for $\mathrm{AICc}(m)$ and (6), under the additional assumption that $\limsup_n r_n < 1$, as well as for $\mathrm{BIC}(m)$ and (7), under the additional assumptions that $\limsup_n r_n < 1$ and $n^{r_n}(1 - r_n)^2 a_n \to 0$, as is easily seen.

To see how AIC, AICc, FPE and BIC perform compared to generalized cross-validation and the $S_p$ criterion, first consider the case where the number of explanatory variables is of the same order as sample size, that is, the case where $|m|/n$ is not close to zero. In that case, (4)–(7) suggest that $\mathrm{AIC}(m)$, $\mathrm{FPE}(m)$, $\mathrm{AICc}(m)$ or $\mathrm{BIC}(m)$ will not give a good estimator for $\rho^2(m)$ or $R^2(m)$. Whenever $|m| > 1$, the expressions (4) and (5) are always smaller than $\rho^2(m)$; hence, $\mathrm{AIC}(m)$ and $\mathrm{FPE}(m)$ tend to underestimate $\rho^2(m)$. Similarly, for $|m| > 1$, the expressions in (6) and (7) are always larger than $\rho^2(m)$, so $\mathrm{AICc}(m)$ and $\mathrm{BIC}(m)$ tend to overestimate $\rho^2(m)$. More importantly, these criteria will typically not select a model with small (conditional or unconditional) mean squared prediction error because the minimizers of $\rho^2(m)$ or $R^2(m)$ over $m \in \mathcal{M}_n$ typically differ from the minimizers of (4), (5), (6) or (7). Hence, if the sample size is small relative to the complexity of the true data-generating process, such that $\rho^2(m)$ is minimized by a model $m$ with $|m|/n$ not close to zero, then the objective functions of AIC, AICc, FPE or BIC can give a distorted picture of that model's performance, both in absolute terms and relative to other candidate models. These model selectors cannot be guaranteed to choose a good model in that situation.

It should be kept in mind that $\mathrm{AIC}(m)$, $\mathrm{AICc}(m)$, $\mathrm{FPE}(m)$ and $\mathrm{BIC}(m)$ do not, in fact, primarily aim to estimate the out-of-sample mean squared prediction error $\rho^2(m)$ (or $R^2(m)$). For example, $\mathrm{AIC}(m)$ is derived from an estimator of the Kullback–Leibler discrepancy between the true and the fitted in-sample predictive distribution; that estimator is asymptotically unbiased, provided model $m$ is correct. Further, $\mathrm{BIC}(m)$ is derived from a first-order expansion of the posterior probability of model $m$ in a Bayesian framework. In certain asymptotic settings where the sample size is typically much larger than the parameters in the model (and for an appropriately chosen class of candidate models), a model minimizing the AIC or the BIC objective function also performs well for prediction out-of-sample in the limit. But, if the number of explanatory variables in the candidate model is not small compared to sample size, this correspondence can break down, as we see here. Similar considerations apply, mutatis mutandis, to $\mathrm{AICc}(m)$ or $\mathrm{FPE}(m)$; see Leeb and Pötscher (2008) for further details.

It remains to consider the case where the number of explanatory variables is of smaller order than sample size. We consider this case for completeness, even though it is not the main focus of this paper. This case is typical for traditional (parametric or nonparametric) large-sample settings, where the sample size is (much) larger than the complexity of the underlying data-generating process so that it can be described by a model that is relatively simple compared to sample size. If $|m|/n$ is small, it is easy to see that the objective functions $\mathrm{GCV}(m)$, $S_p(m)$, $\hat{\rho}^2(m)$, $\mathrm{AIC}(m)$, $\mathrm{AICc}(m)$ and $\mathrm{FPE}(m)$ are essentially equivalent as estimators for $\rho^2(m)$ or $R^2(m)$; the same is also typically true for BIC$(m)$, provided that $|m|/n$ is small enough, in view of (7). In typical parametric settings, this is reflected by the fact that, with probability approaching 1 as sample size increases, these objective functions are minimized by correct models only. However, if $|m|/n$ is small and the simple model $m$ is a good approximation to the true data-generating process, then the noise variance $\sigma^2$ is the dominating factor in both $\rho^2(m)$ and $R^2(m)$. To distinguish between model selection methods here, it is common to consider other performance measures like the (conditional or unconditional) mean of $n(x^{(f)'}\beta - x^{(f)'}\hat{\beta}(\hat{m}))^2$ or variants thereof, where $\hat{m}$ is the model minimizing the objective function under consideration, for example, AIC$(\cdot)$ or BIC$(\cdot)$; see also Remark 2.1. In such a comparison, and in the large-sample limit, BIC is typically found to perform differently from the other model selectors considered here. But, as outlined in the Introduction, the relative efficiency of the post-model-selection estimators obtained by, say, AIC and BIC, respectively, depends crucially on unknown parameters and on sample size, to the extent that either one can be more efficient than the other. We suspect that in such settings, post-model-selection estimators, which can be viewed as 0-1-shrinkage-type estimators, are too crude to perform well in general and that methods based on smooth shrinkage are preferable. This is demonstrated by Goldenshluger and Tsybakov (2003), who propose a smooth shrinkage estimator that is shown to be asymptotically minimax for out-of-sample prediction over Sobolev balls.

## 4. Numerical results

In this section, we investigate the performance of model selectors in finite samples by simulation, where we consider the Gaussian case, as well as several non-Gaussian cases. We focus on 'hard' problems, where the number of parameters is large compared to sample size. We stress that these examples are meant for the purposes of demonstration and should not be mistaken for an exhaustive finite-sample simulation analysis. The simulation results are shown in Figures 2–4, and are explained in the following subsection. For each of three different scenarios introduced below, we consider *one fixed realization of X and Y* (the set of explanatory variables and the response vector, resp.). Given a collection of candidate models that will be chosen later, we compare the estimated performance of each model $m$, that is, GCV$(m)$, with its actual performance, that is, $\rho^2(m)$; see the solid black curve and the solid gray curve, respectively, in Figures 2–4. In addition, the figures also show how AIC$(m)$, AICc$(m)$, FPE$(m)$ and BIC$(m)$ evaluate the models. We have repeated the simulations for other realizations of $X$ and $Y$; the results were essentially unchanged. (The R-code used for the simulations is available from the author on request, together with the results of additional simulation runs.)

### 4.1. Three simulation scenarios

In each of the three scenarios, the explanatory variables $x_j$, $j \geq 1$, and the error $u$ in (1) are taken as mutually independent with mean zero and variance 1 so that $\Sigma$ is the identity and $\sigma^2 = 1$ here. For the actual distribution of the explanatory variables and the error, we consider three distributions – normal, exponential and Bernoulli – each scaled and centered to have mean zero and variance 1. We consider each of these three distributions for the explanatory variables and for the error, resulting in a total of nine combinations, for example, the $x_j$'s are i.i.d. normal and $u$ is normal in (1), the $x_j$'s are i.i.d. (recentered and rescaled) exponential and $u$ is normal in (1), etc. The case where all random variables in (1) are Gaussian has been analyzed in Section 3 from a theoretical perspective. The (recentered and rescaled) exponential and Bernoulli distributions are considered because they are very different from the Gaussian, that is, highly non-symmetric and discrete, respectively. Our simulation results are essentially unaffected by departures from normality. In particular, the results from each of the nine combinations of distributions are visually indistinguishable from each other in graphs like Figures 2–4. In these figures, we therefore only report the results for the case where the explanatory variables are i.i.d. (rescaled and recentered) exponentials and where the error is standard normal. (The results for the other eight combinations are available from the author on request.)

We now describe the three scenarios underlying Figures 2–4; the scenarios differ in the sample size, in the class of candidate models considered and in the underlying regression parameter. For each scenario, we choose the parameters so that the problem is hard, in the sense that there is a rather large number of acceptable models (i.e., models $m$ such that $\rho^2(m)$ is close to $\min_{m \in \mathcal{M}_n} \rho^2(m)$) and in the sense that the acceptable models are rather complex.

For the results in Figure 2, the sample size is $n = 700$ and we consider leading-term submodels, that is, all models $m$ of the form $m = (1, \ldots, 1, 0, \ldots)$ with $|m| = 0, \ldots, 600$; this gives a collection of 601 candidate models. The first 600 coefficients of $\beta$ (in absolute value) are depicted in the top panel of Figure 1; the remaining coefficients of $\beta$ are set equal to zero. The parameter $\beta$ is such that the 'signal-to-noise' ratio $(\mathrm{Var}_{n,\beta,\sigma,\Sigma}[y] - \sigma^2)^{1/2}/\sigma$ equals five; the same also applies to the parameters chosen for Figures 3 and 4. (If the 'signal-to-noise' ratio is chosen too small, only very parsimonious models perform well; increasing the 'signal-to-noise' ratio has the opposite effect. Consistent with the focus of this paper, we have chosen a 'signal-to-noise' ratio between these two extremes.) For Figure 2, the parameter $\beta$ is chosen such that its first 600 components are arranged in 'approximately decreasing' order, while the remaining components are zero. This scenario is meant to reflect a situation where some prior knowledge is available that allows one to arrange the coefficients of $\beta$ by decreasing importance such that the consideration of leading-term submodels is appropriate. Because such prior knowledge is typically incomplete, the coefficients are only approximately ordered (in absolute value) here. The results of this simulation are summarized by the black and gray curves in Figure 2. Black curves depend only on the data like, for example, $\mathrm{GCV}(m)$, while gray curves also depend on the parameters $\beta$, $\Sigma$ and $\sigma$, like, for example, $\rho^2(m)$. The black curves show $\mathrm{GCV}(m)$, $\mathrm{AIC}(m)$, $\mathrm{AICc}(m)$, $\mathrm{FPE}(m)$ and $\mathrm{BIC}(m)$ for each of the 601 candidate models $m$ ordered by $|m|$. For better readability, the points are joined by lines. The minimum of each of these black curves is indicated by a solid dot with the name of the objective function next to it. The black curves have corresponding gray curves.

The gray curves corresponding to GCV($m$), AIC($m$), FPE($m$), AICc($m$) and BIC($m$) are given by $\rho^2(m)$ and by the expressions in (4)–(7), respectively. For reference, the coefficients of $\beta$ (in absolute value) are also plotted at the bottom of Figure 2, with a separate axis on the right.

For the second scenario, which is shown in Figure 3, we take $n = 1300$ and the parameter $\beta$ is such that only its first 1000 components are non-zero. The non-zero coefficients of $\beta$ are 'sparse', in the sense that most of them are rather small (but non-zero), while a few groups of adjacent coefficients are large (cf. the middle panel in Figure 1). Here, we choose a collection of candidate models that can pick-out the few groups of large coefficients. We divide the first 1000 coefficients of $\beta$ into 20 consecutive blocks of length 50 each and consider all candidate models that include or exclude a block at a time, resulting in $2^{20}$ candidate models. With more than a million candidate models, we do not compute GCV($m$) for each model under consideration. Instead, we search through model space using the obvious greedy general-to-specific strategy: fit the 'overall' model containing all 20 blocks and eliminate that block whose elimination leads to the smallest increase in the residual sum of squares (this results in a model containing 19 blocks); now, proceed inductively until all blocks have been eliminated. This procedure gives a data-driven sequence of 20 models of increasing complexity and a corresponding data-driven blockwise rearrangement of the coefficients of $\beta$. (The investigation of alternative search strategies that are potentially superior to the greedy general-to-specific approach is beyond the scope of this paper.) The middle panel of Figure 1 shows the coefficients of $\beta$ (in absolute value) in their original order. At the bottom of Figure 3, the coefficients are rearranged as described above. The description of the curves is as for Figure 2.

For Figure 4, that is, the third scenario, we consider exactly the same setting as for Figure 3, the only exception being that the coefficients of $\beta$ are here not 'sparse' (see the bottom panel in Figure 1). This exemplifies a situation where the collection of candidate models is inadequate for the (unknown) regression parameter.

## 4.2. Discussion

In the setting of Figure 2, the approximations developed in Sections 3.2 and 3.3 for the Gaussian case have clearly 'kicked in': GCV($m$) is very close to the conditional mean squared prediction error $\rho^2(m)$, uniformly over the class of candidate models. Also, AIC($m$), FPE($m$), AICc($m$) and BIC($m$) are close to the quantities in (4)–(7), respectively. Only GCV($m$) gives an accurate indication of the models' performance; the other objective functions do not properly reflect the (relative) performance of the various candidate models. The model minimizing GCV($m$) is very close to the model minimizing the conditional mean squared prediction error $\rho^2(m)$ and the performance of that model is well approximated by the generalized cross-validation criterion. Also, the model minimizing AICc($m$) performs well. This, however, is more of a coincidence than a feature, as it is very easy to find a scenario where AICc($m$) does not perform well; see Figure 4.

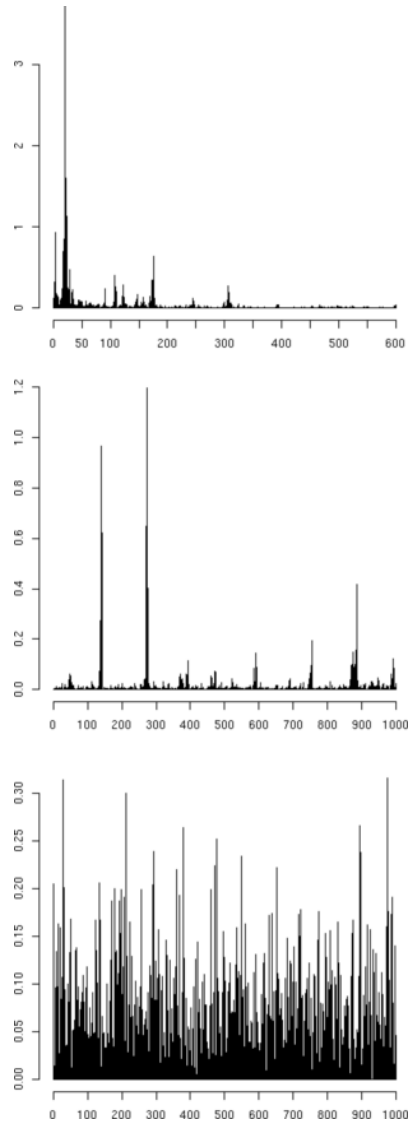

**Figure 1.** Starting from the top, the panels show the absolute values of the non-zero coefficients of the regression parameter $\beta$ used for the simulation results in Figures 2, 3 and 4, respectively. In each case, the parameters are such that the the 'signal-to-noise' ratio is five, that is, $(\mathrm{Var}_{n,\beta,\sigma,\Sigma}[y] - \sigma^2)^{1/2}/\sigma = 5$ with $\sigma = 1$.

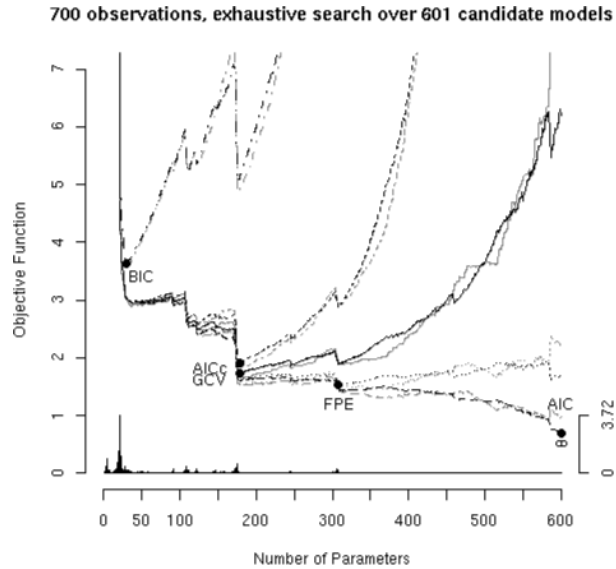

**Figure 2.** Results for the first simulation example. The black curves show GCV($m$) (solid), AIC($m$) (long dashed), FPE($m$) (dotted), AICc($m$) (short dashed) and BIC($m$) (dot-dashed); the minimum of each of these curves is indicated by a black dot with the name of the model selector next to it. The gray curves show $\rho^2(m)$ (solid), as well as the expressions in (4)–(7) (long dashed, dotted, short dashed and dot-dashed, resp.).

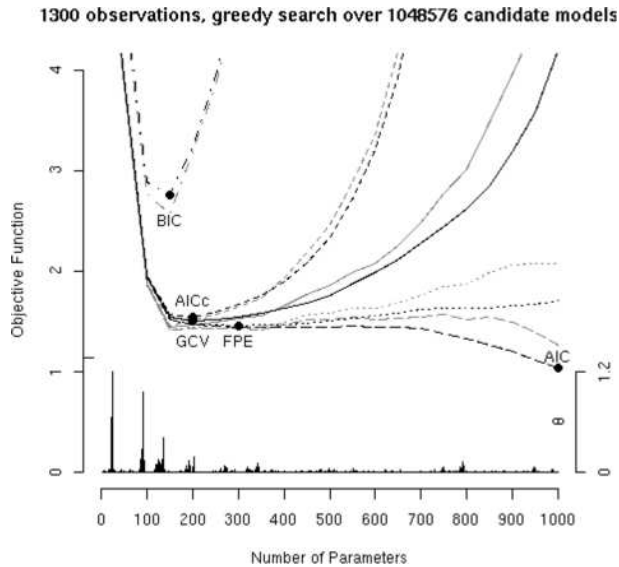

**Figure 3.** Results for the second simulation example. Definition of curves as in Figure 2.

In the settings of Figures 3 and 4, $GCV(m)$ still provides a reasonably good approximation to $\rho^2(m)$, but the approximation is less accurate and $GCV(m)$ tends to underestimate $\rho^2(m)$ for the more complex candidate models. That $GCV(m)$ is less accurate as an approximation to $\rho^2(m)$ is due to the fact that the number of candidate models is much larger here than in the setting of Figure 2. In particular, the number of candidate models is three orders of magnitude larger than the sample size here. (Recall that by partitioning the coefficients of $\beta$ into blocks of length 50, we obtain $2^{20}$ candidate models. Decreasing the block size results in even more candidate models and in deteriorating accuracy; increasing the block size has the opposite effect.) The phenomenon that $GCV(m)$ tends to be smaller than $\rho^2(m)$ is caused by the nature of the greedy search through model space which, in each step, eliminates that block of parameters that results in the smallest increase of the residual sum of squares.

The results in Figure 3 show that $GCV(m)$ continues to perform reasonably well, even if the number of candidate models is much larger than sample size. Again, $GCV(m)$ is close to $\rho^2(m)$ and the model minimizing $GCV(m)$ performs similarly to the overall best candidate model, the minimizer of $\rho^2(m)$. And, as before, the other objective functions do not properly reflect the models' performance. Here, it happens that the models minimizing $BIC(m)$ and $AICc(m)$ also perform well, but, again, this need not be the case in general (BIC performs poorly in Figures 2 and 4, and AICc performs poorly in Figure 4). The model minimizing $GCV(m)$ by using the greedy general-to-specific search through model space performs remarkably well here. For comparison, consider the following (infeasible) procedure: reorder the coefficients of $\beta$ such that their absolute val-ues are decreasing and reorder the columns of $X$ accordingly; after reordering, consider leading-term submodels similarly to the setting of Figure 2 and choose the model for which the conditional mean squared prediction error is minimized. The performance of that model is indicated by the unlabeled extra tick mark on the vertical axis of Figure 3. The performance of the (feasible) procedure that minimizes $\mathrm{GCV}(m)$ by a greedy search is remarkably close to that of the infeasible method just described.

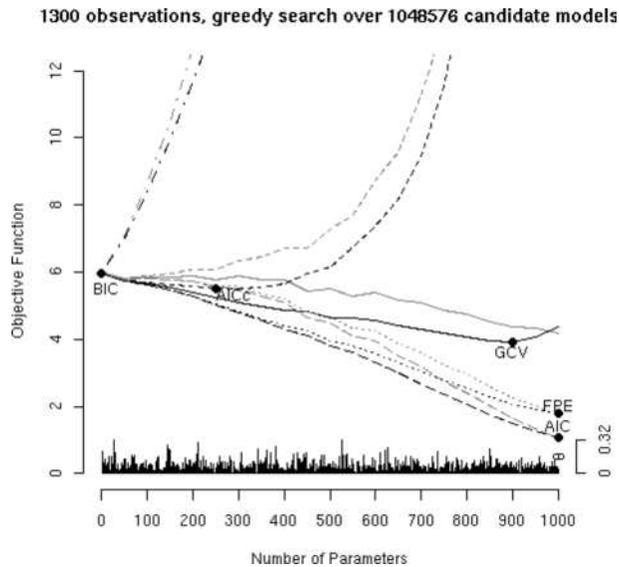

**Figure 4.** Results for the third simulation example. Definition of curves as in Figure 2.

In Figure 4, the largest model with all 1000 coefficients performs best (in terms of conditional mean squared prediction error). This is a situation where the unknown parameter is such that none of the lower-dimensional models perform well. Here, the models minimizing $\mathrm{AIC}(m)$ and $\mathrm{FPE}(m)$ perform very well and the model minimizing $\mathrm{GCV}(m)$ performs comparably, but slightly worse. As before, only $\mathrm{GCV}(m)$ gives a reasonable indication of the models' actual performance, while the other objective functions do not. The models minimizing $\mathrm{BIC}(m)$ and $\mathrm{AICc}(m)$ do not perform well here.

It is striking that the results in Figures 2–4 are basically unaffected by the underlying distribution of the explanatory variables and of the error term in (1). For each of the nine combinations of distributions for the explanatory variables and for the error that we considered, the results are visually indistinguishable from those shown in Figures 2–4. Although our theoretical analysis in Section 3 applies only to the Gaussian case, our simulation results suggest that our main findings are hardly affected by departures from normality, at least in the examples considered here.

# Appendix: Proofs

## A.1. Auxiliary results

The first two lemmas in this section are derived using Chernoff's method, or variants thereof.

**Lemma A.1.** *Let $A$ and $B$ be independent random variables distributed as $\chi_a^2$ and $\chi_b^2$, respectively, with $a, b \in \mathbb{N}$. For each $\varepsilon > 0$, we then have*

$$P\left(\frac{A}{B} - \frac{a}{b} > \varepsilon\right) \leq \mathrm{e}^{-(b/2)\mathcal{K}(a/b,\varepsilon)}$$

*and*

$$P\left(\frac{A}{B} - \frac{a}{b} < -\varepsilon\right) \leq \begin{cases} \mathrm{e}^{-(b/2)\mathcal{K}(a/b,-\varepsilon)}, & \text{if } \varepsilon < a/b, \\ 0, & \text{otherwise.} \end{cases}$$

*The function $\mathcal{K}(\cdot, \cdot)$ is given by*

$$\mathcal{K}(r,c) = (1+r)\log\frac{1+r+c}{1+r} - r\log\frac{r+c}{r}$$

*for $r > 0$ and $c > -r$.*

**Proof.** For $0 < t < 1/2$, we have

$$P(A/B - a/b > \varepsilon) = P(A > B(\varepsilon + a/b))$$
$$= P(\exp(tA) > \exp(tB(\varepsilon + a/b)))$$
$$= E[P(\exp(tA) > \exp(tB(\varepsilon + a/b))||B)]$$
$$\leq E[\exp(-tB(\varepsilon + a/b))(1 - 2t)^{-a/2}]$$
$$= (1 + 2t(\varepsilon + a/b))^{-b/2}(1 - 2t)^{-a/2},$$

where the inequality is based on Markov's inequality, the moment generating function of the $\chi^2$-distribution and the fact that $A$ and $B$ are independent. Rewrite the above inequality as

$$P(A/B - a/b > \varepsilon) \leq e^{-(b/2)f(t)}$$

with $f(t) = \log(1 + 2t(\varepsilon + a/b)) + (a/b)\log(1 - 2t)$. It is elementary to verify that $f(t)$ is maximized over $0 < t < 1/2$ at $t_* = (1/2)\varepsilon/((\varepsilon + a/b)(1 + a/b))$ and that $f(t_*) = \mathcal{K}(a/b, \varepsilon)$. (Note that $f(\cdot)$ is twice continuously differentiable on $(0, 1/2)$; solving $f'(t) = 0$ gives $t_* \in (0, 1/2)$, as above. Because $f''(\cdot)$ is negative, i.e., because $f(\cdot)$ is concave, on $(0, 1/2)$, $f(\cdot)$ attains its maximum at $t_*$.) This gives the first inequality.

The second inequality is trivial in the case $a/b \leq \varepsilon$. For the case $a/b > \varepsilon$, we have

$$P(A/B - a/b < -\varepsilon) = P(B(a/b - \varepsilon) > A)$$
$$= P(\exp(tB(a/b - \varepsilon)) > \exp(tA))$$
$$\leq (1 - 2t(a/b - \varepsilon))^{-b/2}(1 + 2t)^{-a/2}$$

for each $t$ satisfying $0 < t < 1/(2(a/b - \varepsilon))$, by a similar argument as used above. Again, write the inequality in the above display as $P(A/B - a/b < -\varepsilon) \leq \exp(-(b/2)g(t))$ and note that $g(t)$ is maximized at $t_\star = (1/(2(a/b - \varepsilon)))(\varepsilon/(a/b + 1))$ which satisfies $0 < t_\star < 1/(2(a/b - \varepsilon))$, and that $g(t_\star) = \mathcal{K}(a/b, -\varepsilon)$. $\qquad\square$

**Lemma A.2.** *Let $B$ be distributed as $\chi_b^2$ with $b \in \mathbb{N}$. For each $\varepsilon > 0$, we then have*

$$P\left(\frac{B}{b} - 1 > \varepsilon\right) \leq e^{-(b/2)\mathcal{L}(\varepsilon)}$$

*and*

$$P\left(\frac{B}{b} - 1 < -\varepsilon\right) \leq \begin{cases} e^{-(b/2)\mathcal{L}(-\varepsilon)}, & \text{if } \varepsilon < 1, \\ 0, & \text{otherwise.} \end{cases}$$

*The function $\mathcal{L}(\cdot)$ is given by*

$$\mathcal{L}(c) = c - \log(1 + c)$$

*for $c > -1$.*

**Proof.** For the first inequality, fix $t$ satisfying $0 < t < 1/2$ and note that

$$
\begin{aligned}
P(B/b - 1 > \varepsilon) &= P(B > b(1 + \varepsilon)) \\
&= P(\exp(tB) > \exp(tb(1 + \varepsilon))) \\
&\leq e^{-tb(1+\varepsilon)}(1 - 2t)^{-b/2}
\end{aligned}
$$

(as in the proof of Lemma A.1, we use Markov's inequality and the moment generating function of $B$ here). The inequality in the above display can be written as $P(B/b - 1 > \varepsilon) \leq \exp(-(b/2)f(t))$ with $f(t) = 2t(1 + \varepsilon) + \log(1 - 2t)$. The function $f(\cdot)$ is maximized over $(0, 1/2)$ at $t_* = (\varepsilon/2)/(1 + \varepsilon)$ because $f'(t_*) = 0$ and $f''(t) < 0$ for $0 < t < 1/2$. Observing that $f(t_*)$ equals $\mathcal{L}(\varepsilon)$ gives the first inequality.

For the second inequality, assume that $\varepsilon < 1$ (the other case being trivial). An argument similar to that used in the preceding paragraph gives that $P(B/b - 1 < -\varepsilon) \leq \exp(-(b/2)g(t))$ with $g(t) = -2t(1 - \varepsilon) + \log(1 + 2t)$ for $t > 0$. It is elementary to verify that $g(\cdot)$ is maximized for $t > 0$ at $t_* = \varepsilon/(2(1 - \varepsilon))$ and that $g(t_*) = \mathcal{L}(-\varepsilon)$. □

Lemmas A.1 and A.2 give finite-sample analogs to well-known large deviation results. In particular, a result of Killeen, Hettmansperger and Sievers (1972) entails, in the notation of Lemma A.1, that

$$
\frac{1}{b} \log P\left(\frac{A}{B} - \frac{a}{b} \geq \epsilon\right) \xrightarrow[a/b \to r]{b \to \infty} -\frac{1}{2}\mathcal{K}(r, \epsilon).
$$

(In the above relation, $b$ is required to go to infinity and $a/b$ is required to converge to a limit $r \in (0, \infty)$. That relation follows from Example 5.1 of Killeen, Hettmansperger and Sievers (1972) upon expressing $A/B - a/b$ as a linear function of an $F$-distributed random variable.) In finite samples, the first upper bound in Lemma A.1 gives

$$
\frac{1}{b} \log P\left(\frac{A}{B} - \frac{a}{b} \geq \epsilon\right) \leq -\frac{1}{2}\mathcal{K}(a/b, \epsilon).
$$

Similar considerations apply, mutatis mutandis, to the upper bounds given in Lemma A.2. (In view of Theorem 1 of Chernoff (1952), that is obvious because of the way these upper bounds are constructed.)

**Lemma A.3.** *Fix $r > 0$ and consider the functions $\mathcal{K}(\cdot, \cdot)$ and $\mathcal{L}(\cdot)$ defined in Lemmas A.1 and A.2, respectively.*

(i) *For $c$ satisfying $0 \leq c < r$, we have*

$$
\mathcal{K}(r, c) \leq \mathcal{K}(r, -c);
$$

*moreover, for $c$ satisfying $0 \leq c < 1$, the above relation continues to hold with $\mathcal{L}(\cdot)$ replacing $\mathcal{K}(r, \cdot)$.*

(ii)  *For each $c \geq 0$, the functions $\mathcal{K}(r,c)$ and $\mathcal{L}(c)$ are related by*

$$\mathcal{L}\left(\frac{c}{r+1+c}\right) \leq \mathcal{K}(r,c).$$

(iii)  *The function $\mathcal{L}(\cdot)$ is increasing on $[0,\infty)$; for $c$ satisfying $0 \leq c < 1$, $\mathcal{L}(c)$ satisfies*

$$\frac{c^2}{4} \leq \mathcal{L}(c).$$

**Proof.** For part (i), assume first that $0 \leq c < r$. We need to show that $\mathcal{K}(r,c) \leq \mathcal{K}(r,-c)$ or, equivalently, that

$$(1+r)\log\frac{1+r+c}{1+r-c} \leq r\log\frac{r+c}{r-c}. \tag{8}$$

Setting $f(u,v) = u\log((u+v)/(u-v))$, the relation in (8) is equivalent to $f(r+1,c) \leq f(r,c)$. Clearly, this relation is satisfied for $c=0$. With this, it suffices to show that $\partial f(r+1,v)/\partial v \leq \partial f(r,v)/\partial v$ for $0 < v \leq c$. Now,

$$\frac{\partial f(u,v)}{\partial v} = 2\frac{u^2}{(u+v)(u-v)}.$$

To derive (8), it remains to observe that $\partial f(u,v)/\partial v$ is decreasing in $u$ for $r \leq u \leq r+1$ because

$$\frac{\partial^2 f(u,v)}{\partial v\,\partial u} = -4\frac{uv^2}{(u^2-v^2)^2} < 0$$

for $0 < v \leq c$, $c < r$ and $r \leq u \leq r+1$.

To complete the proof of part (i), assume that $0 \leq c < 1$. We need to show that $\mathcal{L}(c) \leq \mathcal{L}(-c)$ or, equivalently, that

$$2c + \log\frac{1-c}{1+c} \leq 0.$$

Write $g(c)$ for the expression on the left-hand side of the above inequality. Clearly, $g(0) \leq 0$. That $g(c) \leq 0$ also holds for $0 < c < 1$ follows upon observing that $g'(c) = -2c^2/(1-c^2)$ is negative for $0 < c < 1$.

For part (ii), write $\mathcal{K}(r,c)$ as $\mathcal{K}(r,c) = h(r+1) - h(r)$ with $h(r) = r\log((r+c)/r)$. It is elementary to verify that $h(\cdot)$ is increasing and concave on $[0,\infty)$: for $r > 0$ and $c \geq 0$, we have

$$h'(r) = \log(1+c/r) - \frac{c}{r+c}$$

$$= -\left(\log\left(1 - \frac{c}{r+c}\right) + \frac{c}{r+c}\right) \geq 0$$

and

$$h''(r) = -\frac{c^2}{r(c+r)^2} \le 0,$$

and this entails that $\mathcal{K}(r,c) \ge h'(r+1) = \mathcal{L}(-c/(r+1+c)) \ge \mathcal{L}(c/(r+1+c))$, where the last inequality follows from part (i).

For part (iii), observing that $\mathcal{L}'(c) = 1 - 1/(1+c)$ shows that $\mathcal{L}(\cdot)$ is increasing on $[0,\infty)$. The lower bound for $\mathcal{L}(c)$ is trivial in the case $c = 0$. For $0 < c < 1$, the lower bound follows upon observing that $\mathcal{L}'(c) \ge c/2$ because $c/2 - \mathcal{L}'(c) = c(c-1)/(2(c+1))$ is negative for such $c$. $\qquad\square$

## A.2. Proofs of main results

**Proof of Proposition 3.1.** In the case where $m$ is of the form $m = (1, \ldots, 1, 0, \ldots)$, the statement in (i) is equivalent to Breiman and Freedman (**1983**), Theorem 1.3 (provided that the quantity $p$ in that theorem is set to $p = |m|$; also, note that the quantity $M_{n,p}$ defined in Breiman and Freedman (**1983**) then coincides with the conditional mean squared prediction error $\rho^2(m)$ considered here). For general $m$ (with $|m| < n$), note that reordering the explanatory variables (and reordering the components of $\beta$ conformably) does not change the conditional mean squared prediction error. Hence, Breiman and Freedman (**1983**), Theorem 3.1 also gives the distribution of $\rho^2(m)$ for general $m$.

The following preliminary consideration is required to derive the second part of the proposition. Throughout the following, fix a candidate model $m \in \mathcal{M}$. Recall the linear model (1) and write $z$ for the $|m|$-vector of those explanatory variables $x_j$ that are included in the model $m$. Because $y$ and $z$ are jointly Gaussian, the conditional distribution of $y$ given $z$ is again a Gaussian. Because both $y$ and $z$ have mean zero, the conditional mean of $y$ given $z$ is a linear function of $z$. Recalling that the conditional variance of $y$ given $z$ is $\sigma^2(m)$, we see that $y\|z \sim N(z'\theta, \sigma^2(m))$ for an appropriate $|m|$-vector $\theta$. In other words, (1) can be rewritten as

$$y = z'\theta + v \qquad (9)$$

with $v \sim N(0, \sigma^2(m))$ independent of $z$.

To prove the statement in (ii), write $Z$ for the $n \times |m|$ matrix of those explanatory variables in the sample that are included in the model $m$. Conditional on $Z$, it follows from (9) and the attending discussion that $\mathrm{RSS}(m)$, that is, the residual sum of squares from regressing $Y$ on $Z$, is distributed as $\sigma^2(m)\chi^2_{n-|m|}$. Because this conditional distribution does not depend on $Z$, the unconditional distribution of $\mathrm{RSS}(m)$ coincides with the conditional distribution. $\qquad\square$

The proof of our main result, that is, Theorem 3.2, rests on the following proposition, which gives an upper bound for $P_{n,\beta,\sigma,\Sigma}(|\hat{\rho}^2(m) - \rho^2(m)| > \epsilon)$ that is tighter, but more complex, than the bound given in Theorem 3.2.

**Proposition A.4.** *In the setting of Theorem 3.2 and for each $\epsilon > 0$, the probability $P_{n,\beta,\sigma,\Sigma}(|\hat{\rho}^2(m) - \rho^2(m)| > \epsilon)$ is not larger than $B_1 + B_2 + B_3 + B_4$. Here, the quantities $B_1$ and $B_2$ are defined as*

$$B_1 = \exp\left[-\frac{n+1-|m|}{2}\mathcal{K}(|m|/(n+1-|m|), \epsilon/(2\sigma^2(m)))\right],$$

$$B_2 = \exp\left[-\frac{n-|m|}{2}\mathcal{L}((\epsilon/(2\sigma^2(m)))(n+1-|m|)/(n+1))\right]$$

*in the case $\sigma^2(m) > 0$ and as $B_1 = B_2 = 0$ otherwise, where the functions $\mathcal{K}(\cdot, \cdot)$ and $\mathcal{L}(\cdot)$ are as in Lemmas A.1 and A.2, respectively. The quantity $B_3$ is set equal to zero in the case $\epsilon/(2\sigma^2(m)) \geq |m|/(n+1-|m|)$; otherwise, $B_3$ is defined as $B_1$ with $-\epsilon$ replacing $\epsilon$. Finally, the quantity $B_4$ is set equal to zero in the case $\epsilon/(2\sigma^2(m)) \geq (n+1)/(n+1-|m|)$; otherwise, $B_4$ is defined as $B_2$ with $-\epsilon$ replacing $\epsilon$.*

**Proof.** In the case $\sigma^2(m) = 0$, both $\rho^2(m)$ and $\hat{\rho}^2(m)$ are equal to zero with probability 1, in view of Proposition 3.1. Hence, the statement of the proposition is trivial in that case. Assume, now, that $\sigma^2(m) > 0$. The probability of interest, that is, $P_{n,\beta,\sigma,\Sigma}(|\rho^2(m) - \hat{\rho}^2(m)| > \epsilon)$, is bounded from above by

$$P_{n,\beta,\sigma,\Sigma}\left(\left|\rho^2(m) - \sigma^2(m)\frac{n+1}{n-|m|+1}\right| > \epsilon/2\right)$$
$$+ P_{n,\beta,\sigma,\Sigma}\left(\left|\sigma^2(m)\frac{n+1}{n-|m|+1} - \hat{\rho}^2(m)\right| > \epsilon/2\right). \tag{10}$$

Let $E$, $F$ and $G$ denote independent, $\chi^2$-distributed random variables with $|m|$, $n - |m| + 1$ and $n - |m|$ degrees of freedom, respectively. Using Proposition 3.1, the probabilities in (10) can be reexpressed in terms of $E$, $F$ and $G$. Simplifying the resulting expressions, we see that (10) is equal to

$$P\left(\left|\frac{E}{F} - \frac{|m|}{n-|m|+1}\right| > \frac{\epsilon}{2\sigma^2(m)}\right)$$
$$+ P\left(\left|\frac{G}{n-|m|} - 1\right| > \frac{\epsilon}{2\sigma^2(m)}\frac{n-|m|+1}{n+1}\right). \tag{11}$$

To complete the proof, we need to show that (11) is not larger than $B_1 + B_2 + B_3 + B_4$. The first term in (11) can be bounded from above using Lemma A.1. In particular, using that lemma with $E$, $F$, $|m|$, $n + 1 - |m|$ and $\epsilon/(2\sigma^2(m))$ replacing $A$, $B$, $a$, $b$ and $\epsilon$, respectively, we see that the first term in (11) is bounded from above by $B_1 + B_3$. For the second term in (11), we use Lemma A.2 with $G$, $n - |m|$ and $(\epsilon/(2\sigma^2(m)))((n-|m|+1)/(n+1))$ replacing $B$, $b$ and $\epsilon$, respectively, and obtain that the second term in (11) is bounded by $B_2 + B_4$. □

**Proof of Theorem 3.2.** Because the case $\sigma^2(m) = 0$ is trivial, we assume that $\sigma^2(m) > 0$. In view of Proposition A.4, it suffices to show that $B_1 + B_2 + B_3 + B_4$ is not larger than the upper bound given by Theorem 3.2.

First, consider the sum of $B_1$ and $B_3$. By Lemma A.3(i), $B_1 + B_3$ is bounded by $2B_1$. Set $r^* = |m|/(n+1-|m|)$ and $c^* = \epsilon/(2\sigma^2(m))$ so that $2B_1 = 2\exp[-((n+1-|m|)/2)\mathcal{K}(r^*, c^*)]$. Now, using Lemma A.3(ii) with $r^*$ and $c^*$ replacing $r$ and $c$, respectively, we see that $2B_1$, and hence also $B_1 + B_3$, is bounded by

$$2\exp\left[-\frac{n+1-|m|}{2}\mathcal{L}\left(\frac{c^*}{r^*+1+c^*}\right)\right].$$

The lower bound for $\mathcal{L}(\cdot)$ provided by Lemma A.3(iii) entails an upper bound for the expression in the preceding display. Simplifying the resulting bound and recalling that $\Psi(\cdot)$ was defined by $\Psi(x) = (x/(x+1))^2/8$, we see that

$$B_1 + B_3 \leq 2\exp\left[-(n+1-|m|)\Psi\left(\frac{\epsilon}{2\sigma^2(m)}\left(1 - \frac{|m|}{n+1}\right)\right)\right].$$

Note that the right-hand side of the above inequality increases if $n+1$ is replaced by $n$.

For the sum of $B_2$ and $B_4$, Lemma A.3(i) shows that $B_2 + B_4$ is bounded by $2B_2$ or, more explicitly, by

$$2\exp\left[-\frac{n-|m|}{2}\mathcal{L}\left(\frac{\epsilon}{2\sigma^2(m)}\left(1 - \frac{|m|}{n+1}\right)\right)\right].$$

Again using Lemma A.3(iii), we get

$$B_2 + B_4 \leq 2\exp\left[-(n-|m|)\Psi\left(\frac{\epsilon}{2\sigma^2(m)}\left(1 - \frac{|m|}{n+1}\right)\right)\right].$$

The upper bounds for $B_1 + B_3$ and $B_2 + B_4$ obtained above immediately entail the upper bound given in Theorem 3.2, completing the proof. $\square$

The following result gives upper bounds for $P_{n,\beta,\sigma,\Sigma}(|S_p(m) - R^2(m)| > \epsilon)$ that parallel those for $P_{n,\beta,\sigma,\Sigma}(|\hat{\rho}^2(m) - \rho^2(m)| > \epsilon)$ given in Proposition A.4 and Theorem 3.2.

**Proposition A.5.** *In the setting of Theorem 3.2 and for each $\epsilon > 0$, the probability $P_{n,\beta,\sigma,\Sigma}(|S_p(m) - R^2(m)| > \epsilon)$ is not larger than $C_1 + C_2$. Here, $C_1$ is defined as*

$$C_1 = \exp\left[-\frac{n-|m|}{2}\mathcal{L}((\epsilon/\sigma^2(m))(n-1-|m|)/(n-1))\right]$$

*in the case $\sigma^2(m) > 0$ and as $C_1 = 0$ otherwise. The quantity $C_2$ is set equal to zero in the case $\epsilon/\sigma^2(m) \geq (n-1)/(n-1-|m|)$; otherwise, $C_2$ is defined as $C_1$, but with $-\epsilon$ replacing $\epsilon$. The upper bound $C_1 + C_2$ is, furthermore, bounded from above by*

$$2\exp\left[-(n-|m|)\Psi\left(\frac{\epsilon}{\sigma^2(m)}\left(1 - \frac{|m|}{n-1}\right)\right)\right],$$

*where $\Psi(\cdot)$ is as in Theorem 3.2.*

**Proof.** As in the proof of Proposition A.4, the case $\sigma^2(m) = 0$ is trivial and we assume that $\sigma^2(m) > 0$. Using Proposition 3.1 and the formulas for $S_p(m)$ and $R^2(m)$ given in Sections 2 and 3, respectively, we see that the probability of interest, that is, $P_{n,\beta,\sigma,\Sigma}(|S_p(m) - R^2(m)| > \epsilon)$, can be written as

$$P\left(\left|\frac{G}{n - |m|} - 1\right| > \frac{\epsilon}{\sigma^2(m)}\frac{n - |m| - 1}{n - 1}\right),$$

where $G$ denotes a random variable that is $\chi^2$-distributed with $n - |m|$ degrees of freedom. Let $\epsilon^* = 2\epsilon(n+1)(n - |m| - 1)/((n-1)(n - |m| + 1))$. Then the expression in the preceding display coincides with the second term in (11) if, in that second term, $\epsilon$ is replaced by $\epsilon^*$. In the proof of Proposition A.4, we have seen that the second term in (11) is bounded by $B_2 + B_4$ (where $B_i$, $i = 2, 4$, are as in Proposition A.4). Using the formulas for $B_2$ and $B_4$ with $\epsilon^*$ replacing $\epsilon$, we obtain the upper bound $C_1 + C_2$. To complete the proof, we use the upper bound for $B_2 + B_4$ obtained in the proof of Theorem 3.2, replace $\epsilon$ by $\epsilon^*$ and simplify. □

**Proof of Corollary 3.3.** The result follows upon noting that

$$P_{n,\beta,\sigma,\Sigma}\left(\sup_{m \in \mathcal{M}_n} |\hat{\rho}^2(m) - \rho^2(m)| > \epsilon\right)$$

$$\leq \sum_{m \in \mathcal{M}_n} P_{n,\beta,\sigma,\Sigma}(|\hat{\rho}^2(m) - \rho^2(m)| > \epsilon) \tag{A.12}$$

$$\leq 4\#\mathcal{M}_n \exp\left[-n(1 - r_n)\Psi\left(\frac{\epsilon}{2c}(1 - r_n)\right)\right].$$

Here, the first inequality is Bonferroni's inequality; the second inequality follows from Theorem 3.2 upon noting that the upper bound in that theorem increases in $|m|/n \leq r_n$ and in $\sigma^2(m) \leq \mathrm{Var}_{\beta,\sigma,\Sigma}[y] \leq c$. □

**Proof of Theorem 3.4.** The plan of the proof is as follows. We first show that the result holds with $\hat{\rho}^2(m)$ replacing $\mathrm{GCV}(m)$ and then that it holds with $S_p(m)$ and $R^2(m)$ replacing $\mathrm{GCV}(m)$ and $\rho^2(m)$, respectively. Finally, we show that $\sup_{m \in \mathcal{M}_n} |\hat{\rho}^2(m) - \mathrm{GCV}(m)|$ and $\sup_{m \in \mathcal{M}_n} |\hat{\rho}^2(m) - S_p(m)|$ are both $O_p(a_n)$, uniformly over the set of parameters where $\mathrm{Var}_{\beta,\sigma,\Sigma}[y] \leq c$. For later use, we note that $a_n \to 0$ implies that $n(1 - r_n)^k \to \infty$ for $k \in \{0, 1, 2, 3\}$ because $a_n^2 \geq \log 2/(n(1 - r_n)^3) \geq \log 2/(n(1 - r_n)^k)$.

To show that (3) holds with $\hat{\rho}^2(m)$ replacing $\mathrm{GCV}(m)$, assume that $\beta$, $\sigma$ and $\Sigma$ satisfy $\mathrm{Var}_{\beta,\sigma,\Sigma}[y] \leq c$ and fix $K > 0$ for the moment. By Corollary 3.3, we see that $P_{n,\theta,\sigma,\Sigma}(\sup_{m \in \mathcal{M}_n} |\hat{\rho}^2(m) - \rho^2(m)| > a_n K)$ is bounded from above by

$$4\exp[-n(1 - r_n)\Psi(Ka_n(1 - r_n)/2c) + \log(\#\mathcal{M}_n)].$$

Because $\Psi(x) = (x/(x+1))^2/8$, the exponent in the preceding display simplifies to

$$-\frac{K^2}{32c^2}\frac{\log(\#M+1)}{(Ka_n(1-r_n)/2c+1)^2} + \log(\#\mathcal{M}_n)$$

$$\leq -\log(\#\mathcal{M}_n+1)\left(\frac{K^2}{128c^2}-1\right),$$

where the inequality holds for sufficiently large $n$, that is, $n \geq n(K)$; here, $n(K)$ is chosen such that $(Ka_n(1-r_n)/2c+1) \leq 2$ for $n \geq n(K)$ (that such $n(K)$ exists follows from $a_n \to 0$). Hence, $\limsup_n \sup_{\beta,\sigma,\Sigma} P_{n,\theta,\sigma,\Sigma}(\sup_{m \in \mathcal{M}_n} |\hat{\rho}^2(m)-\rho^2(m)| > a_n K)$ can be made arbitrarily small by choosing $K$ sufficiently large, where the supremum is taken over all $\beta$, $\sigma$ and $\Sigma$ satisfying $\mathrm{Var}_{\beta,\sigma,\Sigma}[y] \leq c$. This shows that (3) holds with $\hat{\rho}^2(m)$ replacing $\mathrm{GCV}(m)$.

That (3) holds with $S_p(m)$ and $R^2(m)$ replacing $\hat{\rho}^2(m)$ and $\rho^2(m)$, respectively, follows from an argument similar to that used in the preceding paragraph, now using Proposition A.5 and Bonferroni's inequality instead of Corollary 3.3.

We next show that $\sup_{m \in \mathcal{M}_n} |\hat{\rho}^2(m) - \mathrm{GCV}(m)| = O_p(a_n)$, uniformly over the indicated set of parameters; this will entail (3). Let $G$ be a random variable that is $\chi^2$-distributed with $n-|m|$ degrees of freedom, let $\beta$, $\sigma$ and $\Sigma$ be such that $\mathrm{Var}_{\beta,\sigma,\Sigma}[y] \leq c$ and fix $K > 0$ for the moment. We then have

$$P_{n,\beta,\sigma,\Sigma}(|\hat{\rho}^2(m) - \mathrm{GCV}(m)| > a_n K)$$

$$= P\left(\frac{G}{n-|m|} - 1 > \frac{a_n K}{\sigma^2(m)}\frac{(n-|m|+1)(n-|m|)}{|m|} - 1\right),$$

in view of Proposition 3.1(ii) and the fact that both $\hat{\rho}^2(m)$ and $GCV(m)$ are linear functions of $\mathrm{RSS}(m)$. Because $|m|/n \leq 1$ and $\sigma^2(m) \leq \mathrm{Var}_{\beta,\sigma,\Sigma}[y] \leq c$, the expression in the above display is bounded from above by

$$P\left(\frac{G}{n-|m|} - 1 > \frac{a_n K}{c}n(1-r_n)^2 - 1\right)$$

$$= P\left(\frac{G}{n-|m|} - 1 > \frac{K}{c}\sqrt{\log(\mathcal{M}_n+1)n(1-r_n)} - 1\right) \leq P\left(\frac{G}{n-|m|} - 1 > 1\right),$$

where the equality follows by plugging in the formula for $a_n$, and where the inequality holds for sufficiently large $n$, that is, $n \geq n(K)$; existence of such $n(K)$ follows from $\log(\mathcal{M}_n + 1) \geq \log(2)$ and from $n(1-r_n) \to \infty$. The probability of interest, that is, $P_{n,\beta,\sigma,\Sigma}(|\hat{\rho}^2(m) - \mathrm{GCV}(m)| > a_n K)$, is thus bounded from above by

$$P\left(\frac{G}{n-|m|} - 1 > 1\right) \leq \mathrm{e}^{-((n-|m|)/2)\mathcal{L}(1)} \leq \mathrm{e}^{-n(1-r_n)\mathcal{L}(1)/2}$$

for $n \geq n(K)$, where the first inequality follows from Lemma A.2. Arguing as in (A.12), this inequality entails that

$$P_{n,\beta,\sigma,\Sigma}\left(\sup_{m \in \mathcal{M}_n} |\hat{\rho}^2(m) - \text{GCV}(m)| > a_n K\right) \leq \exp[-n(1-r_n)\mathcal{L}(1)/2 + \log \#\mathcal{M}_n]$$

for $n \geq n(K)$. The exponent in the upper bound can be written as

$$-n(1-r_n)[\mathcal{L}(1)/2 - (\log \#M_n)/(n(1-r_n))].$$

The expression in the above display goes to $-\infty$ because $n(1-r_n) \to \infty$, $\mathcal{L}(1)/2 > 0$ and $\log \#\mathcal{M}_n/(n(1-r_n)) \leq a_n^2 \to 0$.

Finally, that $\sup_{m \in \mathcal{M}_n} |\hat{\rho}^2(m) - S_p(m)| = O_p(a_n)$, uniformly over the indicated set of parameters, is established by arguing as in the preceding paragraph, but now using $S_p(m)$ in place of $\text{GCV}(m)$. $\quad\square$

**Proof of Corollary 3.5.** To derive part (i), note that $\rho^2(\hat{m}_n^*) - \rho^2(m_n^*)$ is bounded from below by zero and from above by

$$[\rho^2(\hat{m}_n^*) - \text{GCV}(\hat{m}_n^*)] + [\text{GCV}(\hat{m}_n^*) - \text{GCV}(m_n^*)] + [\text{GCV}(m_n^*) - \rho^2(m_n^*)].$$

By Theorem 3.4, the first and the last term in the above display are $O_p(a_n)$, uniformly over the set of parameters satisfying $\text{Var}_{\beta,\sigma,\Sigma}[y] \leq c$. Because the middle term in the above display is non-positive, the statement in part (i) follows. Part (ii) is a direct consequence of Theorem 3.4.

That parts (i) and (ii) continue to hold with $S_p(\cdot)$ or $\hat{\rho}^2(\cdot)$ replacing $\text{GCV}(\cdot)$ and also with $R^2(\cdot)$ replacing $\rho^2(\cdot)$ follows by repeating the argument in the preceding paragraph with the corresponding replacements. $\quad\square$

# Acknowledgements

I am particularly grateful to Andrew Barron, Dietmar Bauer, Manfred Deistler, David Findley, Richard Nickl and Benedikt M. Pötscher for inspiring critique and invaluable feedback. Also, comments from the participants of departmental seminars at the Universities of Maryland, Connecticut, Exeter, Michigan and Vienna, and at Yale University, are greatly appreciated. Finally, I would like to thank Rudy Beran for writing his inspiring papers.